\documentclass[aps,preprint]{revtex4}
\usepackage{graphicx,bm,amssymb,amsmath}

\begin{document}
\preprint{Accepted by Nanotechnology 2008}
\title{Analyzing capacitance-voltage measurements of vertical wrapped-gated
nanowires}

\author{O. Karlstr\"om and A. Wacker}
\affiliation{Mathematical Physics, Lund University, Box 118, 22100 Lund, Sweden}
\author{K. Nilsson, G. Astromskas, S. Roddaro, L. Samuelson, 
and L.-E. Wernersson}
\affiliation{Solid State Physics, Lund University, Box 118, 22100 Lund, Sweden}
\date{August 21, 2008}

\begin{abstract}
The capacitance of arrays of vertical wrapped-gate InAs nanowires are analyzed.
With the help of a Poisson-Schr\"odinger solver, information about the 
doping density can be obtained directly. Further features in
the measured capacitance-voltage characteristics 
can be attributed to the presence of surface states as well
as the coexistence of electrons and holes in the wire. 
For both scenarios, quantitative estimates are provided. 
It is furthermore shown
that the difference between the actual capacitance and the geometrical limit
is quite large, and depends strongly on the nanowire material.
\end{abstract}

\maketitle

\section{Introduction}
Large efforts are currently made to develop new schemes for the manufacturing
of nanowire (NW) devices\cite{AgarwalAP2006}. E.g., 
recent progress in vertically processed semiconductor NWs with a
cylindrically symmetric gate electrode, such as the wrapped insulator-gate
field effect transistor\cite{ThelanderIEDL2008}, have shown great
promise for application in future highly scaled electronic devices. Besides
the possibility of incorporating wrapped gates for improved gate control, NWs
offer other intriguing possibilities not easily accessible to planar designs,
such as an inherent one-dimensionality as well as relaxed constraints in terms
of combining highly lattice-mismatched materials along the NW channel due to
radial relaxation of interface strain\cite{BjorkNano2002,LarssonNT2007}.

\begin{figure}[bt]
\includegraphics[width=0.9\columnwidth]{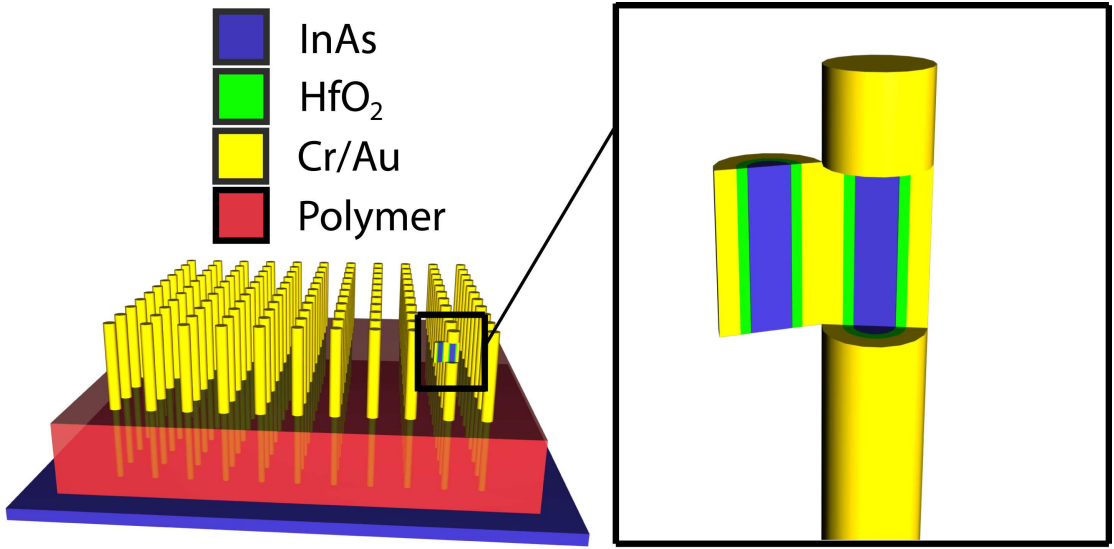}
\caption{Illustration of the array of NWs. Each NW
  consists of an InAs core surrounded by a dielectric layer of $\rm HfO_2$ and
  a Cr/Au gate. The lifting layer of polymer is inserted to reduce the
  capacitance between gate and substrate.}
\label{Fig1}
\end{figure}

In order to facilitate the development of high performance NW electronic
devices, it is important to accurately characterize the doping profile
and interface properties of
the system as well as further fundamental material parameters. 
Si-based MOS-stacks have long benefited from well developed
capacitance-voltage (CV) measurement schemes for device
characterization. However, similar methods have so far been largely
unavailable for NW devices. Recently CV characteristics of individual Ge NWs
on SiO$_{2}$/Si substrates were measured\cite{TuNL2007}. Here we are focusing
on vertical wrapped gate NWs for which we have earlier demonstrated a scalable
processing protocol for routine CV spectroscopy\cite{RoddaroAPL2008}.
An overview of the device structure considered here 
is illustrated in Fig.~\ref{Fig1}. The
device consists of an 11x11 array of InAs NWs, covered with a $10\,{\rm nm}$
conformal $\rm HfO_2$ dielectric layer and a top Cr/Au metalization as one of
the capacitor electrodes. The capacitance is measured between this gate
electrode and the core of the wire, which is connected via the substrate.
Details on the fabrication protocol and measurement techniques can be found
elsewhere\cite{RoddaroAPL2008}. In this
paper we further develop the quantitative as well as qualitative understanding
of these measurements, and also discuss various features of the
CV-characteristics in detail.

\section{The model}

The nanowires are modelled as cylinders where the radius of the InAs will be
referred to as $R_{\mathrm{nw}}$, whereas $R_{\mathrm{gate}}$ is the radius of
the InAs and HfO$_{2}$ together. Typical measures are length $h=680$ nm, and
radial dimensions $R_{\mathrm{nw}}=27$ nm, $R_{\mathrm{gate}}=37$ nm.

We assume that the unintentional 
doping of the NWs is of n-type. By changing the applied gate voltage,
$V_{\mathrm{gate}}$, the charge within the NW, $Q$, can be affected. This
defines the capacitance

\begin{eqnarray}
C=\frac{dQ}{dV_{\mathrm{gate}}}
\end{eqnarray}

The total charge $Q$ is the spatial integral over the charge distribution
$\rho(r)$ containing  three components $\rho=e(N_{D}-n_{e}+n_h)$ inside the
InAs nanowire. Here, $e$ is
the positive elementary charge and  $N_{D}$, $n_{e}$ and $n_h$ are the
concentrations of ionized donors, electrons, and holes, respectively. The
donors are assumed to be entirely ionized at room temperature as considered
throughout this paper. All concentrations are assumed to depend only on the
radial distance $r$ from the centre of the NW. This greatly facilitates the
calculations as it reduces the problem to one dimension. Working with
homogenous doping profiles is motivated by averaging over arrays of long
wires. However, microscopic fluctuations are neglected by this approach.

The effective density of states masses for electrons and holes in InAs are
$m_{e}^{\ast}=0.023m_{e}$ and $m_{h}^{\ast}=0.41m_{e}$
respectively\cite{NakawaskiPHB1995,Kittel1996}. Non-parabolicity effects are
neglected albeit they can cause some quantitative deviations for large
positive biases. For the determination of hole concentrations, we use the
bandgap $E_{g}=0.54$eV for InAs NWs with wurtzite
structure\cite{TragardhJAP2007}. The relative dielectric constant is 
$\varepsilon\approx 15$ both for the InAs NW and the $\rm HfO_2$ insulator 
\cite{Kittel1996,KukliCVD2002,Wheeler2007}.

To calculate the capacitance, a Poisson-Schr\"odinger solver was implemented
similar to the one demonstrated in \cite{WangSSE2006,GnaniSSE2006}. 
For a given charge distribution $\rho(r)$, the cylindrical geometry permits 
the use of Gauss law to solve Poisson's equation for the electrostatic 
potential $\phi(r)$. Thus, we obtain the Hamiltonian
\begin{eqnarray}
H=-\frac{\hbar^{2}}{2}\frac{1}{m_{e}^{\ast}}\nabla^{2}-e\phi(r)
\label{EqHam}
\end{eqnarray}
together with the boundary condition for the wavefunction $\Psi(R_{\mathrm{nw}})=0$
assuming that the $\rm HfO_{2}$ layer constitutes an infinite barrier.
Due to the cylindrical symmetry of the problem,
Bessel functions were used as a basis set for the diagonalisation of
Eq.~(\ref{EqHam}), resulting in the wave functions $\Psi_{\nu j}(r,\varphi)$ 
and corresponding energies $E_{\nu j}$. Here $\nu$ and
$j$ are the angular and radial quantum numbers, respectively.
The relatively long length of the wires
compared to their radial dimensions justifies the use of a 1D density of
states along the z-direction, $g(E_{z})$. Using Fermi-Dirac statistics
the electron concentration is then given by
\begin{equation}
n_e(r)=\sum_{\nu j}|\Psi_{\nu j }(r)|^{2}\int_{0}^{\infty}
\frac{g(E_{z})dE_{z}}{\exp((E_{z}+E_{\nu j})/kT)+1}
\end{equation}
Here the electrochemical potential $E_F$ was set 
to zero in the wire, which is used as a reference point throughout this paper.
Together with the similarly calculated hole concentration $n_h$, this provides
a new charge distribution $\rho(r)$. By iterative solution of the Poisson and
Schr{\"o}dinger equation, a self-consistent solution is obtained.

\begin{figure}
\includegraphics[width=0.9\columnwidth]{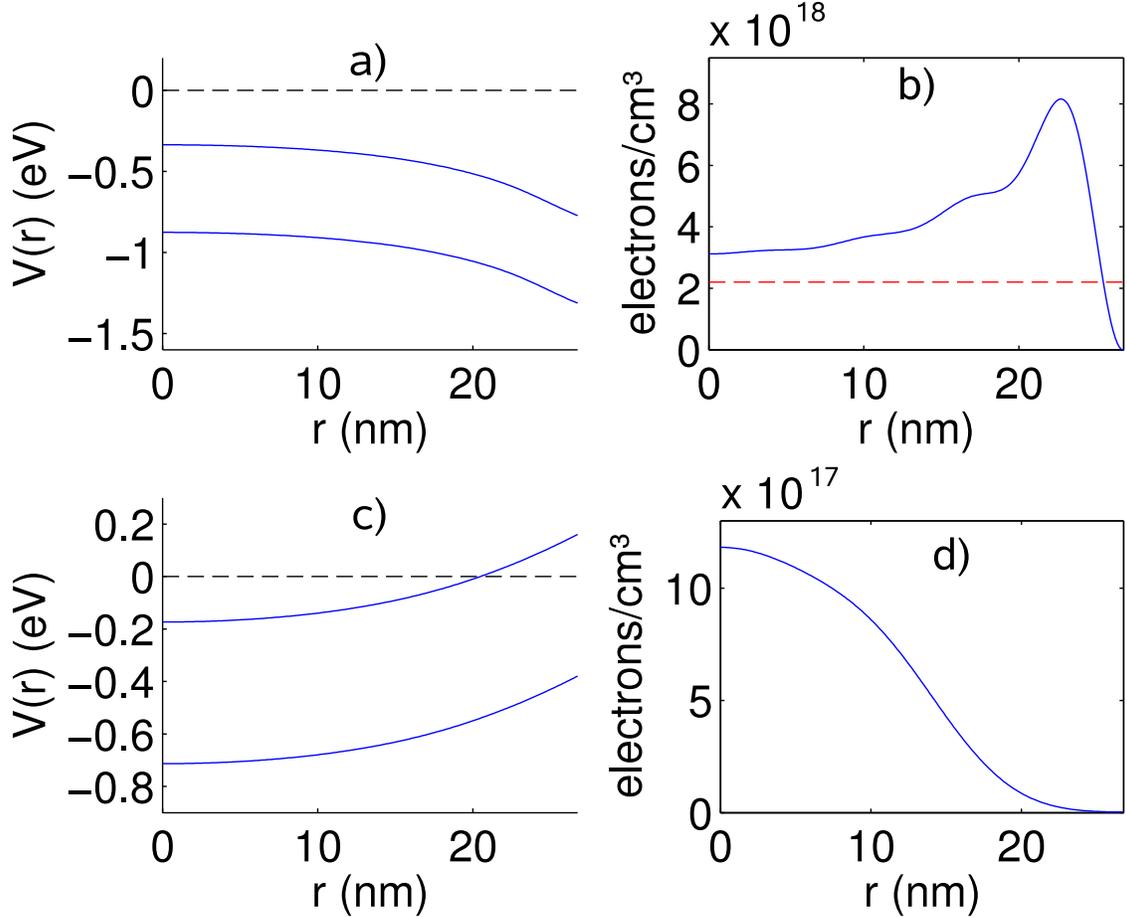}%n=27 n=104
\caption{a) Potential $V(r)=-e\phi(r)$ and valence band edge $V(r)-E_g$
  for $\phi_0(R_{\mathrm{gate}})=1.08$ V, corresponding to the operation point
  1 of Fig.~\ref{Fig3}. The dashed line depicts $E_{F}$ of the wire, which is
  set to zero as reference. b) corresponding electron distribution
  $n_{e}(r)$. c) and d) are the same plots for
  $\phi_0(R_{\mathrm{gate}})=-0.46$ V, (operation point 2). Parameters:
  $N_D=2.2\cdot 10^{18}/\mathrm{cm}^3$ [dashed line in b)], 
  $R_{\mathrm{nw}}=27$ nm.}
\label{Fig2}
\end{figure}

In Fig.~\ref{Fig2} calculated potentials and electron distributions are
shown. Figs.~\ref{Fig2}a) and b) correspond to a positive electrical potential
at the gate electrode. The corresponding bending of the conduction band edge
causes electrons to accumulate close to the semiconductor-oxide interface
(accumulation mode). 
Figs.~\ref{Fig2}c) and d)  correspond to a negative electrical potential
at the gate electrode. Here the electrons are depleted close to the surface of
the wire (depletion mode).

In the simulation we assume that the insulating oxide layer is neutral of
charge and use the corresponding electric  potential
$\phi_0(R_{\mathrm{gate}})$ at the gate as a parameter. However, trapped
charges are likely to occur both at the surface of the wire and within the
oxide\cite{SzeBook1985}.  If an areal charge density $\sigma$ is present
at position $R_{\mathrm{trap}}$ the  potential for $r\ge R_{\mathrm{trap}}$ is
modified according to
\begin{equation}
\phi(r)=\phi_{0}(r)
-\frac{\sigma R_{\mathrm{trap}}}{\varepsilon\varepsilon_{0}}\ln
\left(\frac{r}{R_{\mathrm{trap}}} \right)
\label{EqPhiTrap}
\end{equation}
while $\phi(r)$ is not affected for $r<R_{\mathrm{trap}}$. 
Experimentally, one measures the applied gate
voltage $V_{\mathrm{gate}}$ which is related to the actual electric potential
at the gate via
\begin{equation}
V_{\mathrm{gate}}=\phi(R_{\mathrm{gate}})
+\frac{W_{\mathrm Cr}-\chi_{\mathrm InAsNW}}{e}
\label{EqVgate}
\end{equation}
Here $W_{\rm Cr}$ is the work function of Cr and $\chi_{\rm InAsNW}$
the electron affinity of the InAs in the nanowire.  As long as the trapped
charges are unchanged they provide a  constant bias offset
\begin{equation}
V_{\mathrm{gate}}=\phi_0(R_{\mathrm{gate}})+V_0
\end{equation}
between the measured bias
and the parameter $\phi_0(R_{\mathrm{gate}})$ used in the simulation. As both
the magnitude of the trapped charges and the electron affinity of InAs in the
wurtzite modification appearing in the nanowire are not a priori known, we use
$V_0$ as a fit parameter in the subsequent data analysis. 

\section{Understanding the experimental data}

\begin{figure}
\includegraphics[width=0.9\columnwidth]{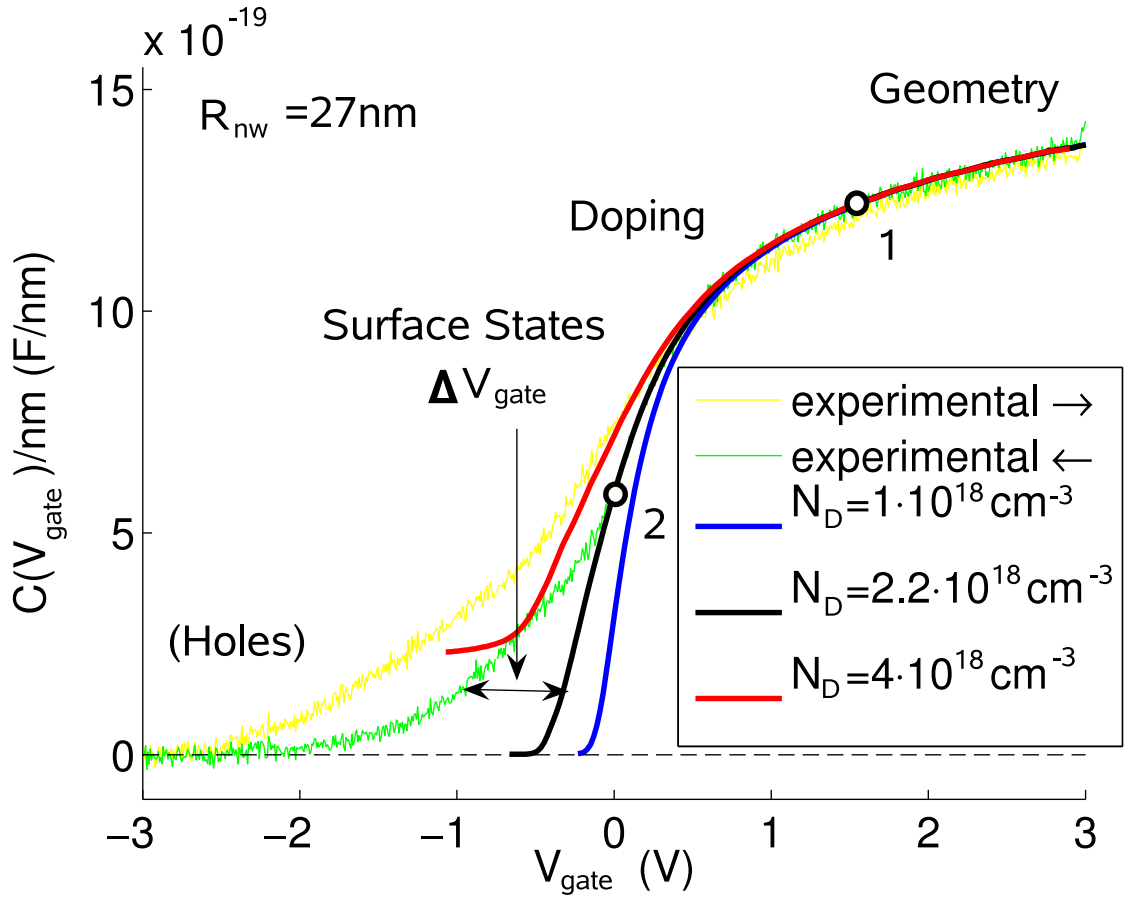}
\caption{Experimentally measured capacitance and
  theoretical fits for a mean wire radius $R_{\mathrm{nw}}=27$ nm. The green
  curve shows the measured capacitance when sweeping from positive to negative
  bias, while the yellow curve is for the opposite sweep direction. The
  gate-substrate capacitance, which is independent of bias, has been
  subtracted.} 
\label{Fig3}
\end{figure}

In Fig.~\ref{Fig3} the experimentally measured capacitance for the device with
mean radius 27 nm is shown together with our calculations for different doping
densities. The main trend is the increase of capacitance with bias. This can
be attributed to the location of charges inside the NW. From Gauss law it
follows that the change in $Q$ induced by a change in bias, will be larger if
$\rho(r)$ is affected at large values of $r$. As can be seen in
Fig.~\ref{Fig2}b) (corresponding to the operation point 1 in  Fig.~\ref{Fig3}),
the charges are mainly located close to the NW surface in the 
accumulation region.
In contrast, they are close to the centre of the wire at operation point 2, see
Fig.~\ref{Fig2}d). This explains, why the capacitance is higher at point 1
than at point 2.

The measurements reported here are performed by adding a small
AC-signal with a frequency of 20 MHz atop the
DC gate-bias. The impedance is found to be almost
completely imaginary showing a negligible influence of any series
resistance to the accumulation capacitance. For positive bias
the measured capacitance shows less than 10 \% frequency
dispersion between 50 kHz and 100 MHz demonstrating good properties
of the metal-oxide-semiconductor structure and that a 
reliable quantitative data analysis can be performed for $V_{\textrm{gate}}>0$.
In contrast, for $V_{\textrm{gate}}<0$
the frequency dispersion is more pronounced, 
which we attribute to the presence of surface states,
see Sec.~\ref{SecSurfaceStates}.

The calculations provide a capacitance per wire length, as given in
Fig.~\ref{Fig3}. For the experimentally measured capacitance $C$, the gate
substrate capacitance, which is independent of bias, see \cite{RoddaroAPL2008}, has been subtracted and the result has been
divided by an effective length $L$, 
which has been taken as a second fit parameter
in addition to $V_0$. The data  shown assumes an effective length
corresponding to an effective number of 90
wires with the nominal length of 680 nm.  This has to be compared with a total
of 121 wires in the fabricated array.  The difference can be attributed to
lacking electrical connection of some wires and/or variations from the nominal
length.

Each fabricated array showed a certain distribution of radii
$R_{\mathrm{nw}}$\cite{RoddaroAPL2008}. To accommodate for this, an ensemble
of three different values of $R_{\mathrm{nw}}$ was used in the simulation for
each device, while the oxide thickness of 10 nm was held constant. The
calculated curves in Fig.~\ref{Fig3} and \ref{Fig4} are the results from such
ensembles with appropriate weights for each radius reflecting the measured
radius distribution.\footnote{We used  $R_{\mathrm{nw}}=25\mathrm{ nm}:
0.5\times25\mathrm{ nm}+0.3\times27.5\mathrm{ nm}+0.2\times22\mathrm{ nm}$\\
$R_{\mathrm{nw}}=27\mathrm{ nm}: 0.4\times26\mathrm{ nm}
+0.45\times28.5\mathrm{ nm}+0.15\times24\mathrm{ nm}$\\
$R_{\mathrm{nw}}=28.5\mathrm{ nm}: 0.4\times27.5\mathrm{ nm}
+0.45\times31\mathrm{ nm}+0.15\times25\mathrm{ nm}$\\
$R_{\mathrm{nw}}=30\mathrm{ nm}: 0.4\times30\mathrm{ nm}
+0.4\times32\mathrm{ nm}+0.2\times27.5\mathrm{ nm}$\\ 
$R_{\mathrm{nw}}=31\mathrm{ nm}: 0.35\times30\mathrm{ nm}
+0.5\times32.5\mathrm{ nm}+0.15\times28\mathrm{ nm}$}

In Fig.~\ref{Fig3} it is indicated that different factors, such as geometry,
doping, surface states, and inversion (holes), 
affect the capacitance at different
gate voltages. We will in turn explain the contribution from each factor.
Measurements indicate that the down-sweep curve (green curve) is closer to
equilibrium as time-dependent sweeps are more stable and the up-sweep  (yellow
curve) relaxes towards the down-sweep curve. Therefore, all fits of our
stationary model are made to the down-sweep curve.

\subsection{Geometry}

According to Fig.~\ref{Fig3} the capacitance is independent of the donor
concentration in the accumulation mode, but rather depends on the geometry of
the wire. The reason is that during accumulation electrons are added close to
the semiconductor-oxide interface, see Fig.~\ref{Fig2}b. Thus different doping
concentrations result in the same capacitance. The capacitance in this region
can therefore not be used to determine the doping concentration. Fitting to
the experimental data instead determines the effective length of the wires.

\begin{figure}
\includegraphics[width=0.9\columnwidth]{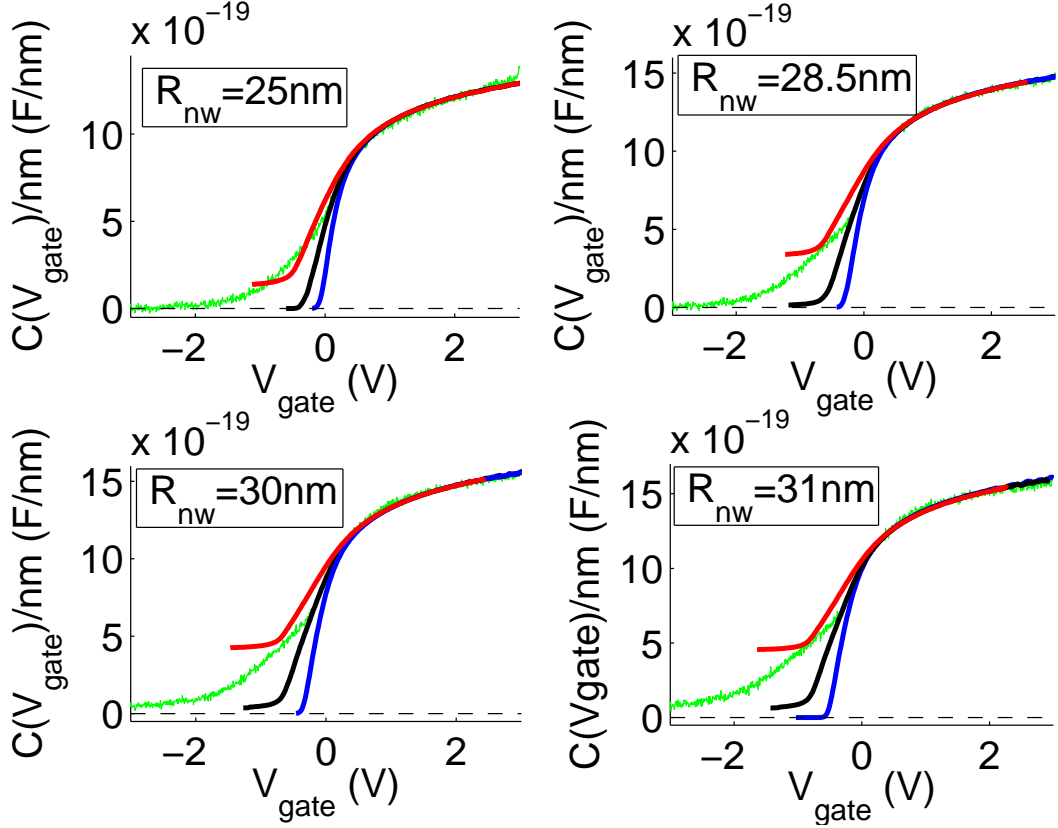}
\caption{Experimentally measured down-sweep capacitance (green)
  and simulation results for different NW radii $R_{\mathrm{nw}}$. The doping
  densities for the different simulations are identical to Fig.~\ref{Fig3}.} 
\label{Fig4}
\end{figure}

In Fig.~\ref{Fig4} the capacitance (experimental down-sweep) for 
NWs with different radii $R_{\mathrm{nw}}$ are shown. All fabricated arrays
have 121 NWs, and the effective number of wires used for fitting $L$
is displayed in Table \ref{TabScale}.
As a general trend, the capacitance increases with the radius 
$R_{\mathrm{nw}}$ as there is a larger surface area of the NWs.
Note that the capacitance in the accumulation region is
significantly lower than the geometrical capacitance
$C/L=2\pi\epsilon\epsilon_0/\ln(R_{\mathrm{gate}}/R_{\mathrm{nw}})$, which is
$2.48 \cdot 10^{-18}$ F/nm for $R_{\mathrm{nw}}=25$ nm and 
$R_{\mathrm{gate}}=35$ nm. This deviation is
strongly dependent on the density of states for the nanowire wire material, 
see Sec.~\ref{SecMaterial}.

\begin{table}
\begin{tabular}{|l|l|l|l|l|}
\hline
$R_{\mathrm{nw}}$ (nm) & \# wires 
&\multicolumn{3}{c|}{offset $V_0$ (V) } \\ 
&&  $N_D=10^{18}/\mathrm{cm}^3$ &
$N_D=2.2\cdot 10^{18}/\mathrm{cm}^3$ &
$N_D=4\cdot 10^{18}/\mathrm{cm}^3$ 
\\ \hline
25 & 71 & 0.27 & 0.42 & 0.63\\ \hline
27 & 90 & 0.26 & 0.39 & 0.60\\ \hline
28.5 & 86 & 0.13 & 0.32 & 0.53 \\ \hline
30 & 90 & 0.12  & 0.28 & 0.51 \\ \hline
31 & 84 & -0.05 & 0.14  & 0.38 \\ \hline
\end{tabular}

\caption{Effective number of wires and bias offset $V_0$ 
  resulting in a best fit of the capacitance-bias
  characteristics for different arrays of NWs and different assumed doping
  densities.}
\label{TabScale}
\end{table}

\subsection{Doping}

As the bias is lowered, the electrons will be depleted in the NW. A large
doping concentration provides a positive background charge with a strong
attraction for the electrons, requiring a higher negative gate voltage to
deplete all the electrons. Thus, the doping concentration strongly affects the
slope of the capacitance curve around $V_{\rm gate}=0$ V. Comparison between
fits and experimental data shows that $N_{D}=4\cdot10^{18}$cm$^{-3}$ is too
flat while $N_{D}=1\cdot10^{18}$cm$^{-3}$ is too steep. The doping of
$N_{D}=2.2\cdot10^{18}$cm$^{-3}$ fits very well up to the "kink" at point 2 of
the experimental down-sweep curve in Fig.~\ref{Fig3}. This also holds for all
devices shown in Fig.~\ref{Fig4}. As all devices were manufactured using the
same method, this provides a consistent estimate of $2.2\cdot10^{18}$cm$^{-3}$
for the doping density.

\subsection{Surface States}
\label{SecSurfaceStates}
The experimental curves exhibit a "kink" at point 2 of
Fig.~\ref{Fig3}, which is most clearly seen for the down-sweep curve.
For biases below the kink the measured capacitance is 
significantly larger than the simulation result. In this
section we attribute this to the emptying of
surface states, situated at the semiconductor-oxide interface.

The surface states are assumed to be located in the forbidden
gap~\cite{DayehAPL2007}, and are treated as
donors~\cite{AffentauscheggSST2001}. As the states are located in the bandgap,
they remain filled as long as $E_{F}$ is in the conduction band. Theoretical
evaluation of the potential at $R_{\mathrm{nw}}$, for the doping of
$N_{D}=2.2\cdot10^{18}$cm$^{-3}$, shows that $E_{F}$ is 0.1-0.2eV below the
conduction band edge at the kink, see Fig.~\ref{Fig2}c). 
This makes it
feasible that the kink is an effect of such surface states. For gate voltages
more negative than the one corresponding to the kink, surface states are
emptied which produces the discrepancy between simulations and experiment.

While conduction band electrons as majority carriers
can almost adiabatically follow the AC signal at 20 MHz,
the filling and emptying of surface states is a slower
process. Therefore these states do not directly contribute 
to the capacitance under our measurement conditions.
For lower AC frequencies, we observe an enhancement of the capacitance
in the bias region below the kink which we attribute to
the direct contribution of surface states. The situation is however intricate
due to a strong variation of charging times and its analysis is a matter of
ongoing research. Thus we focus on the high-frequency data at 20 MHz here.

Even if the surface states do not directly contribute to the 
capacitance 
signal, they affect the capacitance-voltage characteristics
indirectly by the presence of a bias-dependent surface charge.
This positive
charge screens effectively the gate bias from the NW, so that a more negative
$V_\mathrm{gate}$ is needed to deplete the  electrons from the wire. This can
be quantified using  Eqs.~(\ref{EqPhiTrap},\ref{EqVgate}). An
additional (bias-dependent) surface charge density  $\delta\sigma$ at
$R_{\mathrm{nm}}$ gives an additional shift of the gate bias 
$V_{\mathrm{gate}}=\phi_{0}(R_\mathrm{gate})+V_0+\delta V$ with
\begin{equation}
\delta V=-\frac{\delta\sigma R_{\mathrm{nw}}}{\varepsilon\varepsilon_{0}}\ln
\left(\frac{R_\mathrm{gate}}{R_{\mathrm{nw}}} \right)
\label{EqSurfaceEst}
\end{equation}
This explains the shift of the capacitance curve to lower bias while emptying
the surface states. With $\delta V$ being of the order of $-1$ V, 
positive surface charge densities of $\sim 10^{13}/\mathrm{cm}^2$ are 
observed for gate biases far in the depletion region.

To summarize, if surface states are not included in the model, the simulations
should be fitted to the part of the down-sweep curve right of the
kink. The difference in $V_\mathrm{gate}$ between experiment and simulation
below the kink allows for an estimate of the
surface charge density.

\subsection{Holes}

As the electrochemical potential in the wire starts to approach the valence
band, holes will appear. The creation and annihilation of holes are slow
processes. %, typically requiring the interaction with thermal phonons.  
The holes, like the surface states, will therefore not directly contribute to
the capacitance at high frequencies but instead screen the electrons. The high
effective mass of the holes imply a large density of states which results in a
strong screening. If holes are created, practically no electrons will be
removed any longer from the conduction band. The constant electron
concentration causes the capacitance curve to flatten out at a finite
value. This can be seen for the simulated doping
$N_{D}=4\cdot10^{18}$cm$^{-3}$ in Fig.~\ref{Fig3}. The lower doping
concentrations are too small for inversion to start before the last electrons
have been removed, and the capacitance goes to zero in agreement with the
experimental data.

From Fig.~\ref{Fig4} we see that inversion has an effect for the
thicker wires also for $N_{D}=2.2\cdot10^{18}$cm$^{-3}$. This is in agreement
with the measured capacitance which seems to flatten out at a finite value for
negative bias rather than dropping to zero as for the thinner wires.

\begin{figure}
\includegraphics[width=0.9\columnwidth]{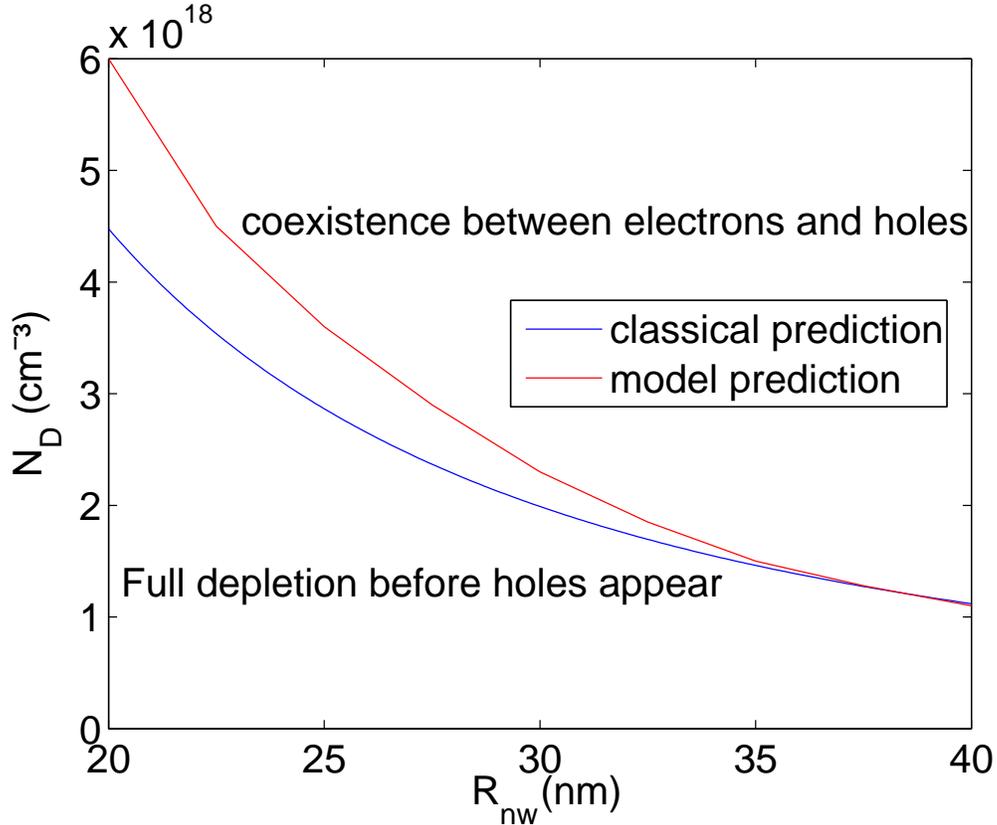}
\caption{Doping density of InAs NWs required for the coexistence of holes and
  electrons} 
\label{FigCoexist}
\end{figure}

In this context it is interesting to evaluate the doping required for the
coexistence of holes and electrons. For zero temperature, this
situation corresponds to the electrochemical potential entering the
valence band at $r=R_{\mathrm{nw}}$ before it leaves the conduction band at
$r=0$. Thus, a certain  doping concentration is required if
the inversion should have any effect on the capacitance.
This doping concentration can be
classically estimated with help of Gauss law 
\begin{eqnarray}
N_{D}^{\rm coexist}(R_{\mathrm{nw}})=\frac{4\varepsilon\varepsilon_0
  E_{g}}{e^{2}R_{\mathrm{nw}}^{2}}
\label{EqCoestEst}
\end{eqnarray}
The required doping as a function of $R_{\mathrm{nw}}$ can also be estimated
with the Poisson-Schr\"odinger solver. The results are displayed in
Fig.~\ref{FigCoexist}. 
Due to the confinement energy, the prediction of the simulation is higher,
than for the classical estimate. The difference is most significant for small
$R_{\mathrm{nw}}$. For large values of $R_{\mathrm{nw}}$ the classical
prediction is equal to, or even larger than the model prediction. The reason
is that the full bandgap is used in Eq.~(\ref{EqCoestEst}). In contrast, due
to the thermal distribution, holes are already occupied slightly before the
electrochemical potential enters the valence band. 
From Fig.~\ref{FigCoexist} we
conclude that at the estimated doping concentration 
$N_{D}=2.2\cdot10^{18}$cm$^{-3}$, inversion will not affect
the capacitance for $R_{\mathrm{nw}} \lesssim 30$ nm in accordance with the
findings of Fig.~\ref{Fig4}.

\section{Comparison between different materials}
\label{SecMaterial}

%GaAs from Sze m*=0.063 eps=12.4

\begin{figure}
\includegraphics[width=0.9\columnwidth]{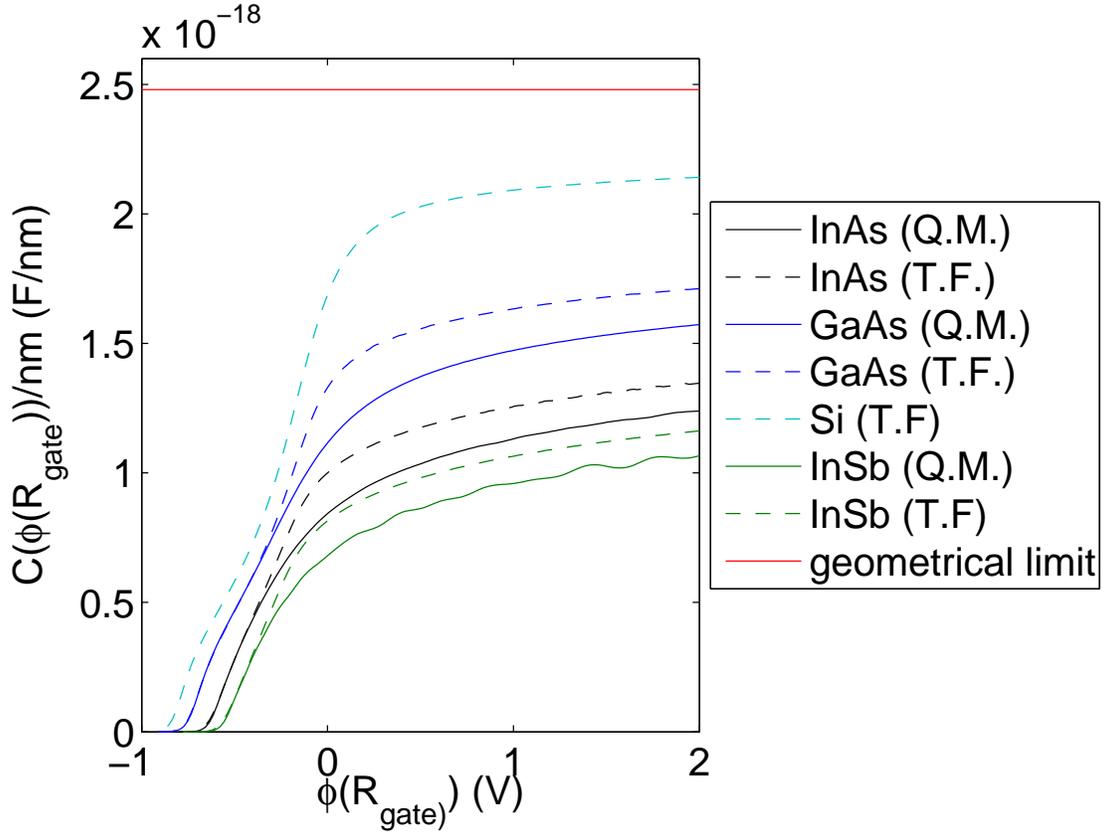}
\caption{Comparison between theoretical capacitance and geometrical limit
  $=2.48\cdot10^{-18}$F/nm, for $R_{\mathrm{nw}}=$25 nm
  $R_{\mathrm{gate}}=$35 nm $N_{D}=2\cdot10^{18}$ /cm$^{3}$, plotted against 
  the electrostatic potential at $R_{\mathrm{gate}}$.} 
\label{FigMaterials}
\end{figure}

We will now show that equating the capacitance of a NW with its geometrical
limit can lead to large overestimates. In Fig.~\ref{FigMaterials} the
theoretically calculated capacitances of InAs-, InSb, GaAs- and Si-NW:s are
plotted. The effective masses of these materials are 0.023, 0.0145,
0.063 and 0.33~\cite{SzeBook1985} respectively. (For Si the
density-of-state mass per conduction band valley was 
used and the valley degeneracy
was explicitly taken into account for.) They are all surrounded by
HfO$_{2}$ and have the same geometry, $R_{\mathrm{nw}}=25$nm and
$R_{\mathrm{gate}}=35$nm, i.e they have the same geometrical limit. For
comparison the results from a Poisson-Thomas-Fermi solver have been
plotted as well. In the Thomas-Fermi approximation, the electron concentration at each
point is determined by the distance between the conduction band edge and
$E_{F}$. Multiplying the Fermi-Dirac distribution with the three-dimensional
density of states and integrating over the conduction band yields the electron
concentration as 
\begin{eqnarray}
n_{e}(r)=\int_{0}^{\infty}\frac{\frac{1}{2\pi^{2}}\left(\frac{2m}{\hbar}\right) ^{3/2}\cdot E^{1/2}dE}{\exp((E-e\phi(r))/(kT))+1}
\end{eqnarray}
where the conduction band edge is given by the electrostatic
potential times the electron charge $-e\phi(r)$. 
{\bf Again we use $E_{F}=0$} as a
reference. 

Due to the complex band structure of Si, only the Poisson-Thomas-Fermi results
have been calculated for this material. The main difference between the two
solvers is that classically the electron concentration does not have to be zero
at the semiconductor-oxide interface, resulting in a higher capacitance. It is
evident that the two models give the same results for a repulsive gate
voltage. In this case the electron concentration approaches zero 
at $R_{\mathrm{nw}}$ as the potential surpasses the electrochemical potential,
see also \cite{WangSSE2006}. 

There is a reason why the geometrical capacitance cannot be reached even
classically. Let us compare with a metallic coaxial cable. In this case the
geometrical limit is reached as all charge is added at the surface. This
requires an infinite density of states, which can not be found in a
semiconductor, and thus the geometrical capacitance will not be
reached. Materials with a high effective mass, such as Si, have a large
density of states and will therefore come closer to the geometrical limit.

Frequently, it is assumed that the capacitance is constant for large positive
gate voltages. Both experimental and theoretical results point out that this
is not the case, the capacitance is slowly increasing with bias. As the gate
becomes more attractive, electrons will be filled higher up in the conduction
band. Since the 3D density of states goes as $E^{1/2}$, the density will
increase as the bias is raised, resulting in a larger capacitance.

\section{Conclusions}

It has been shown how information about the doping and surface state
concentrations can be obtained for semiconductor nanowires by comparing
experimental and simulated capacitance data. The above procedure suggests how
the contribution from these two factors can be separated. Analysis
shows that the doping density $N_{D}=2.2\cdot10^{18}/\mathrm{cm}^{3}$ 
gives good
agreement with the experimental data. The surface state charge at full
depletion is estimated to be around $10^{13}$ $e$ per cm$^{2}$.
Finally, the actual capacitance can be significantly
smaller than its geometrical limit even in accumulation mode. The difference
is of particular importance for nanowires with low effective mass.

%\section*{Acknowledgements}
\acknowledgments
This work was supported by the Swedish Research Council, 
the Swedish Foundation for
Strategic Research, the Knut and Alice Wallenberg Foundation, 
the EU-project NODE 015783, and the Italian Ministry of
University and Research.

\end{document}